\documentclass{sigchi}

\usepackage{balance}  
\usepackage{graphics} 
\usepackage[T1]{fontenc}
\usepackage{txfonts}
\usepackage{mathptmx}
\usepackage[pdftex]{hyperref}
\usepackage{color}
\usepackage{booktabs}
\usepackage{textcomp}
\usepackage{microtype} 
\usepackage{ccicons}  

\usepackage{todonotes}

\def\plaintitle{Spot that Bird: A Location Based Bird Game}
\def\plainauthor{Sneha Mehta}

\def\plainkeywords{Authors' choice; of terms; separated; by
  semicolons; include commas, within terms only; required.}

\makeatletter
\def\url@leostyle{%
  \@ifundefined{selectfont}{
    \def\UrlFont{\sf}
  }{
    \def\UrlFont{\small\bf\ttfamily}
  }}
\makeatother
\def\pprw{8.5in}
\def\pprh{11in}

\definecolor{linkColor}{RGB}{6,125,233}
\hypersetup{%
  pdftitle={\plaintitle},
 pdfauthor={\plainauthor},
  pdfkeywords={\plainkeywords},
  bookmarksnumbered,
  pdfstartview={FitH},
  colorlinks,
  citecolor=black,
  filecolor=black,
  linkcolor=black,
  urlcolor=linkColor,
  breaklinks=true,
}

\begin{document}

\title{\plaintitle}

\numberofauthors{1}
\author{%
  \alignauthor{Sneha Mehta\\
    \affaddr{Department of Computer Science and Center for Human Computer Interaction}\\
    \affaddr{Virginia Tech, Blacksburg, VA, USA}\\
    \email{snehamehta@vt.edu}}\\
}

\maketitle

\begin{abstract}
 In today's age of pervasive computing and social media people make extensive use of technology for communicating, sharing media and learning. Yet while in the outdoors, on a hike or a trail we find ourselves inept of information about the natural world surrounding us. In this paper I present in detail the design and technological considerations required to build a location based mobile application for learning about the avian taxonomy present in the user's surroundings. It is designed to be a game for better engagement and learning. The application makes suggestions for birds likely to be sighted in the vicinity of the user and requires the user to spot those birds and upload a photograph to the system. If spotted correctly the user scores points. I also discuss some design methods and evaluation approaches for the application.
\end{abstract}



\section{Introduction}
Countless studies have shown and proven that regular exercises improve our health, fitness and overall quality of our lives. Hiking is one of the most low stress and fun physical activity that helps reduces the risk of a variety of health issues such as heart diseases, hypertension and diabetes \cite{lloyd2010health}. Hiking usually can be solitary or a in a group, can be in the wilderness or just a plain walk in a city. With all the health benefits and fun factors associated with hiking it is hard to deny
that most hikers have little or no knowledge about the wildlife on their trails. While hiking usually involves people exploring nature trails intertwining through forests full of biodiversity, yet it is a pity that very few tools exist to enhance our knowledge about the habitat we're exploring.
\begin{figure}[ht!]
 \centering
  \includegraphics[width=0.5\textwidth]{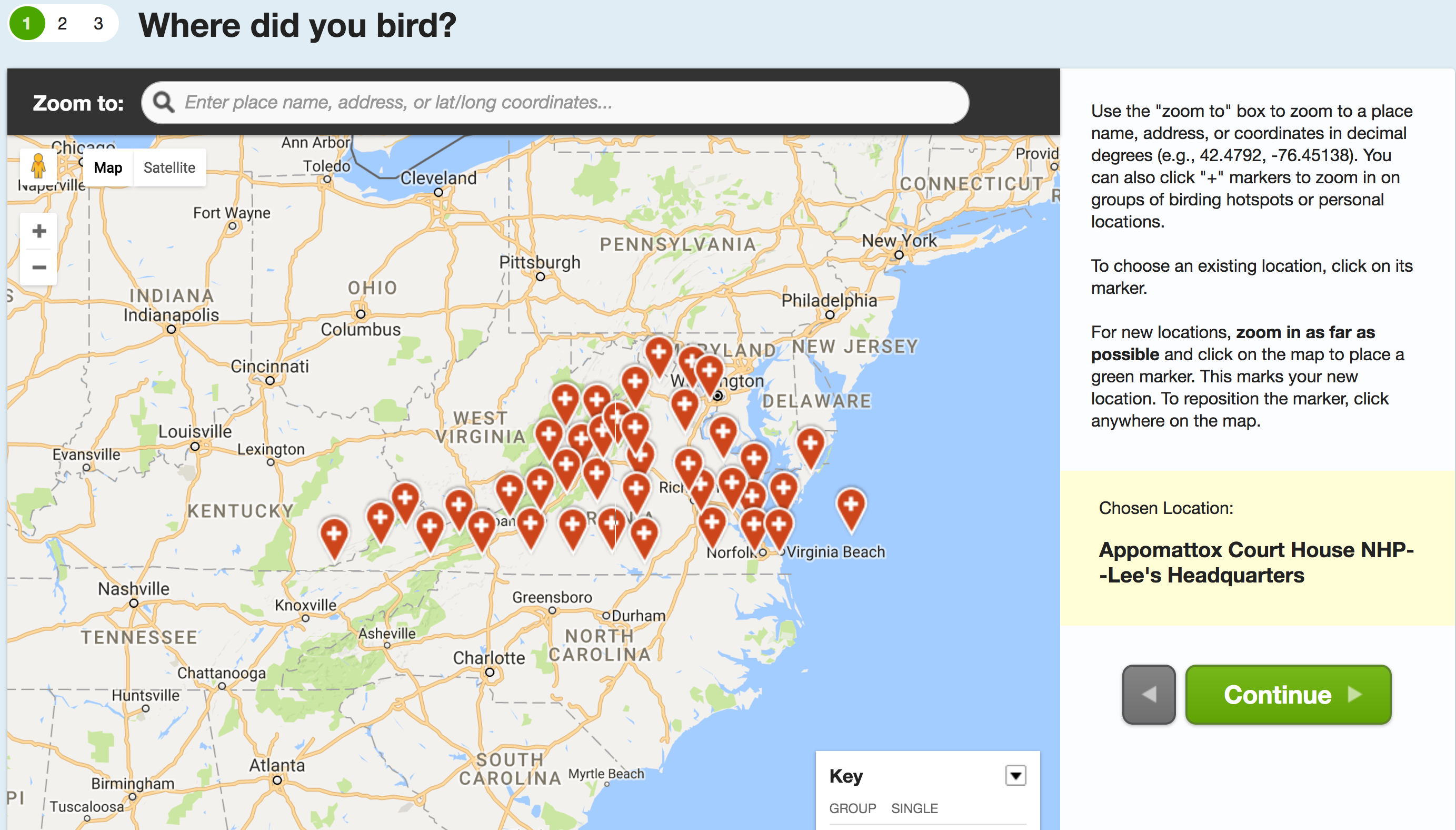}
 \caption{From www.ebird.org. Birding hotspots in the state of Virginia.}
\label{fig:hotspots}
\end{figure}
What are those birds whose ambient sounds you hear while on a trail or those countless trees that you see around you. We wish to be more aware of the things around us while on a hike to gain even more from the experience. Even though there are field guides\footnote{http://merlin.allaboutbirds.org/} and mobile apps\footnote{http://merlin.allaboutbirds.org/} available, using them could be a cognitively demanding activity for hikers aiming for a low stress outdoor activity. 
Some trails also have information boards about the flora and fauna to expect on the hike and the history of the place although many don't. In contrast recent games like Pok\'emon Go \footnote{http://www.pokemongo.com}, Ingress\footnote{https://www.ingress.com/} and Zombies, Run!3\footnote{https://www.zombiesrungame.com/ } that embed game play in physical world have led to anecdotal increase in the physical activity \cite{info:doi/10.2196/jmir.6759}. A natural next step is to integrate traditional learning tools such as bird field guides and mobile games to create a low stress but engaging learning experience. With this motivation I aim to build a mobile game for beginners named Spot that Bird (StB) with the following goal.

\begin{itemize}
\item Novice and non-birders be able to identify and remember names of the common birds in their surroundings.
\end{itemize}
To address the above research goal I propose design and technological considerations in developing such a game. In this game, based on the location of the user, the system generates suggestions of birds likely to be sighted in the user's vicinity. Using the description and other information provided by the application the user has to spot and photograph the birds. The hope is that the process of identifying and photographing birds in the wild with the aid of the application will help them learn about birds. My long term goal is to build an app that facilitates visual learning for a variety of things in the wild such as plants, wildlife, architectural styles etc. Works such as \cite{Belongie201615} are promising efforts in that direction. In this work I mainly focus on birds due to wide availability of data about birds on websites such as ebird \footnote{www.ebird.org} and Flickr.

\section{Related Work}
My work is a location based mobile application and it touches different domains such as computer vision, location based systems and use of mobile devices for learning.

\subsection{Location Based Systems}
With the increasing ubiquity of portable computing devices like mobile phones that come with a variety of sensors such as GPS, camera and accelerometer, applications making using of these sensors have burgeoned. My proposed application uses camera and GPS. In \cite{Raper:2007:CEL:1392056.1392058} authors give a detailed overview of location based technology and discuss the HCI and technical  challenges. Kray et al. \cite{Kray:2003:PRI:604045.604066} performed a comparative analysis of different wayfinding visualization techniques to find that users resonate with the 3D virtual world representation to orient themselves in the real world. Hence it is no surprise that location based applications like Google Maps\footnote{maps.google.com}, Apple Maps or even crowd generated maps like Open Street Maps \footnote{www.openstreetmap.org} that present wayfinding instructions to the users as a 3D map have enjoyed phenomenal success and have become integral part of people's lives. 
In \cite{Puikkonen:2009:TDB:1658550.1658566} authors find that the visualizations and UI designs resembling conventional outdoor maps or floor layouts are not optimal for indoor navigation and present design recommendations for indoor navigation systems. Location based systems have also been used for locating shops or other POIs  \cite{Cheverst:2000:DCE:332040.332047} \cite{Puikkonen:2009:TDB:1658550.1658566}. Dix has a different view regarding map representations as described in his paper Mental geography and Wonky maps In Proc. GeoHCI \cite{Hecht:2013:GHI:2468356.2479637}. Above approaches use an exact representation of the physical space where all locations are reduced to cartesian coordinates. Dix emphasizes the importance of local maps that are created paying more emphasis to human spatial cognition as opposed to cartesian correctness.
Similarly in \cite{Kim:2016:GPS:2858036.2858440} authors Kim et al. present an interactive spatial analogy map creation tool from location data appearing in text.
In \cite{Brule:2016:MMI:2858036.2858375} authors present and evaluate multi-sensory maps to help visually impaired children acquire spatial skills.

\subsection{LBS for social applications}
\cite{ICWSM112751} use gravity models to explain the spatial structure of location based social networks like Foursquare, BrightKite. 
\cite{Ye:2010:LRL:1869790.1869861} investigate research issues in realizing location recommendation services for large-scale location-based social networks, by exploiting the social and geographical characteristics of users and locations/places. While the above papers discuss social location based applications, \cite{Posti:2014:UJH:2598510.2598592} present the new concept of asocial hiking application where the goal is to avoid other hikers while on a hike. The app helps users generate solitary hiking routes based on photographs posted on sites like Flickr. Their mobile application prototype also scans Wi-Fi signals to detect
other hikers nearby and warns of their approach. Like above approaches my application uses the location data of the user's device but is  not designed to be a social(or asocial) application. Above applications use location information of the device to calculate the proximity of other users whereas my application uses the location based data to compute the proximity of birds on a trail.

\subsection{Computer Vision}
While verifying a photo clicked by a user I propose to use a hybrid (computer vision and human) approach \cite{Branson:2010:VRH:1888089.1888123} \cite{Branson2014}. From the perspective of computer vision there are two tasks to be accomplished given a bird photo -- 1) Detect where the bird is in the picture 2) Verify if the bird belongs to the species user clicked the picture for. The latter task which is to verify if the bird in the picture belongs to a specific bird species with possibly subtle differences between other similar looking bird species is related to the problem of fine-grained classification \cite{Simon15:NAC} \cite{Freytag14:BFF}.  Fine-grained classification has been an active challenge in the computer vision community and is challenging mainly because of large inter-class differences and high intra-class similarity. Introduction of datasets such as CUB-200 \cite{WelinderEtal2010} have led to advancements in research in automated recognition of bird species. Researchers have proposed a variety of approaches including creating part models \cite{Simon15:NAC} \cite{Simon14:PDD} or learning mid-level features based on the query image \cite{Freytag14:BFF}. 
Some efforts leverage and complement the competencies of humans and machines to semi-automate the process of recognition. Humans are extremely good at identifying colors or parts such as beak, feather etc. whereas machines are good at remembering complex characterizations of different species of birds. Efforts such as \cite{420} use these complementary skills of humans and machines to propose human-in-the-loop \cite{Cui2016} approaches for fine-grained recognition, where humans have to answer intelligently generated questions by machines like 'Click on the head', 'Is the bill black?'. 
Success of computer vision algorithms largely depends on the quality of the data sets those algorithms are trained on. \cite{Horn2015} use citizen scientists and group of volunteers with interests in specific domains such as birds, insects, architecture etc. to collect high quality data set and build a bird recognition app for recognition North American birds. 

The above approaches are designed to automate the process of bird recognition with only occasional and shallow inputs from humans. My app is designed to educate humans while on a hike or a trail and make them more aware of their surroundings while enjoying the process. As later developed my approach also relies on human feedback for accurate recognition.

One of the main goals of the app is for novice/non-birders to be able to identify and remember common birds in the wild.  In \cite{Bonsignore:2013:SSL:2491500.2491506} authors discuss how storytelling can impact children's learning ability and how mobile devices can be used effectively as learning tools. Authors developed a very popular app for creating and sharing stories in the wild, with the capabilities of integrating media such as images and sound clips in their stories. In this application authors accomplish learning through creating and sharing stories with the help of visual and aural media. Similarly we use bird photographs and bird sounds along with game-play to accomplish the goal of learning.

\section{Gameplay}
The user begins by feeding into the system(the mobile app) a route for an upcoming trail or hike. When on the trail the app notifies the user about common nearby birds. Information about the birds such as their sounds, their photographs and their habits are presented to the user to help them locate the birds. The users have to spot the birds and photograph them using the camera provided in the app. The system then uses the photo and the auxiliary information to verify if the photograph clicked is indeed the picture of the bird suggestion. If it matches the user gets points and advances in the game. As the user advances in the game suggestions for rarer birds are made which are even harder to spot and photograph.

\section{Methods}
Websites like ebird \footnote{www.ebird.org} are a large repository  of bird checklists submitted by users across the world. Along with checklists people also upload a large number of bird pictures including other auxiliary information such as time of the sighting, condition of the sighting, location and count.  

\subsection{Research on User Group}
Following the principals of user centric design I propose to include user feedback to inform my design decisions. User feedback contributes novel development ideas as well as possible further use cases \cite{Posti:2014:UJH:2598510.2598592}. Online surveys including questions with 7 point Likert scale and subjective responses should be conducted. Focus group sessions involving non-birders, novice birders and expert ornithologists should be conducted to identify user hiking behaviors, mobile phone usage in the outdoors, knowledge about bird taxonomy, interest in learning about the natural world and attitudes towards interruptions while hiking. Besides focus groups in person interviews with non-birders, novices and experts should be conducted. 

Ornithologist interviews and possible collaborations will be instrumental in deciding the key factors of the app design, including important factors affecting bird sightings, most distinct bird features and best ways to spot birds. Since spotting birds in the wild is a difficult task app should be carefully designed to make the process user friendly.

During the focus groups and the interviews the design of the app proposed below should be presented to identify ease of use, passive/active notification preference when a bird is in the vicinity(haptic feedback such as vibration or a sound notification), the level of detail to be presented to the novice vs expert users etc. Demographic data should also be collected with respect to the above design considerations to identify the popularity of the app in one demographic vs other. 

\section{System}
In this section I describe in the detail the technology and the design of the application.

\subsection{Generating bird suggestions}
The first important step in developing the app will be downloading data from ebird. ebird is essentially a crowdsourced archive of bird data from around the world. It allows users to upload bird sightings along with other information such as observation data, observation type (traveling, stationary, historical, incidental), time of the sighting, species identified etc. Observation location can be pin-pointed on a google map as a new location or one of the several birding hotspots in the region can be selected.

\begin{figure}[ht!]
 \centering
  \includegraphics[width=0.5\textwidth]{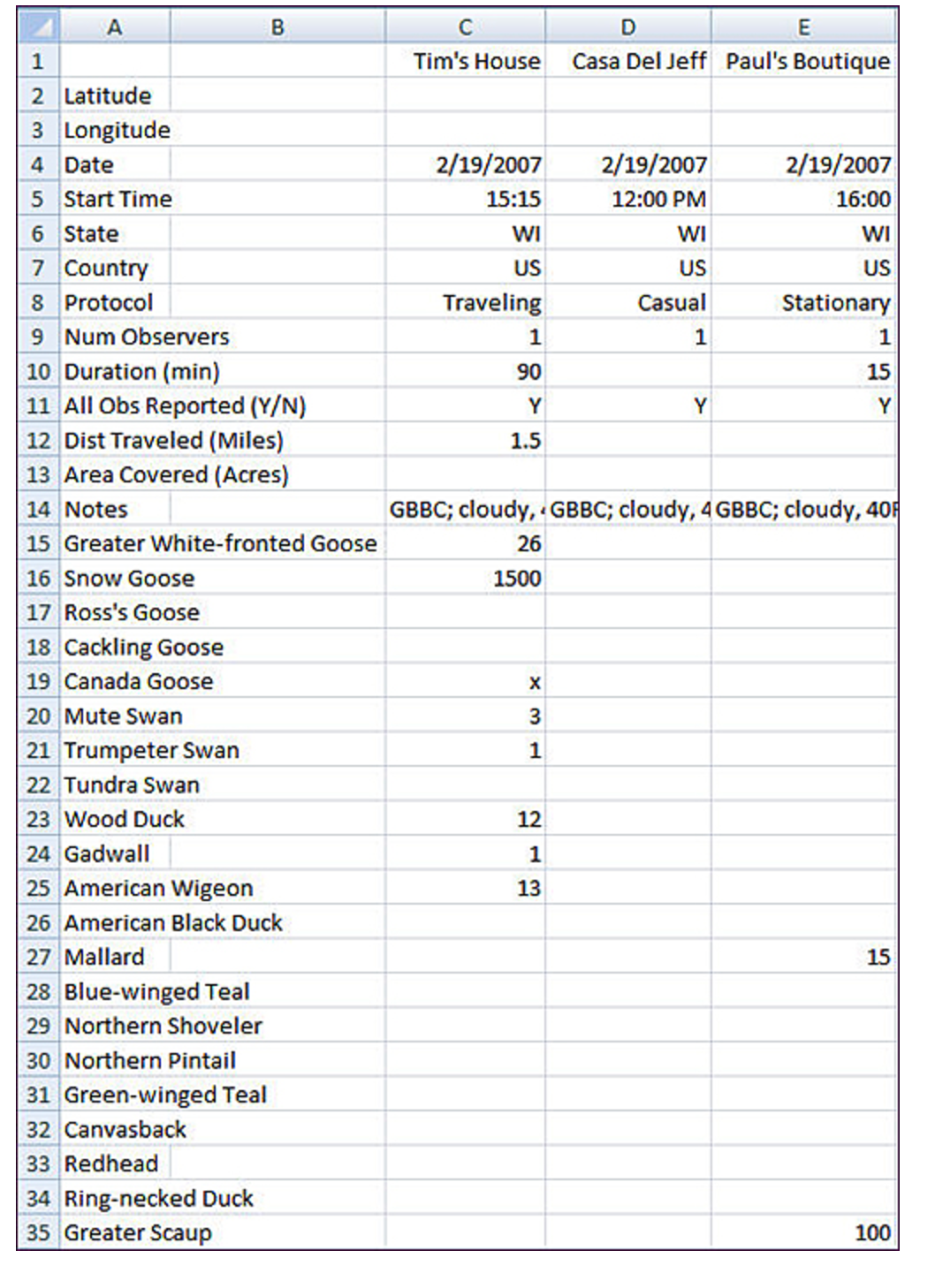}
 \caption{From www.ebird.org. Spreadsheet format for uploading a checklist to ebird. Reporting starts from Row 15 onwards. Row 1 to Row 14 is auxiliary information required.}
 \label{fig:checklist-format}
 \vspace*{0.1in}
\end{figure}

Figure \ref{fig:hotspots} shows the birding hot-spots as recorded in the state of Virginia. The website also lets the users import their bird checklists from an excel sheet. The checklist to be imported has to be in a pre-specified format and rigorously demands information such as  row 1 to row 15 in Figure \ref{fig:checklist-format}. Bird checklists are organized by location generated from the bird sighting data submitted by the users. A very important piece of information in the bird checklist is the bird count.
For a new StB user only when she comes under the proximity of frequently sighted birds should she be notified so that it is easier to cross the levels while still keeping the task challenging. To identify common birds, ornithologists should be consulted as to what needs to be the sighting count for the bird to be classified as a common bird.

\begin{figure}[ht!]
 \centering
  \includegraphics[width=0.5\textwidth]{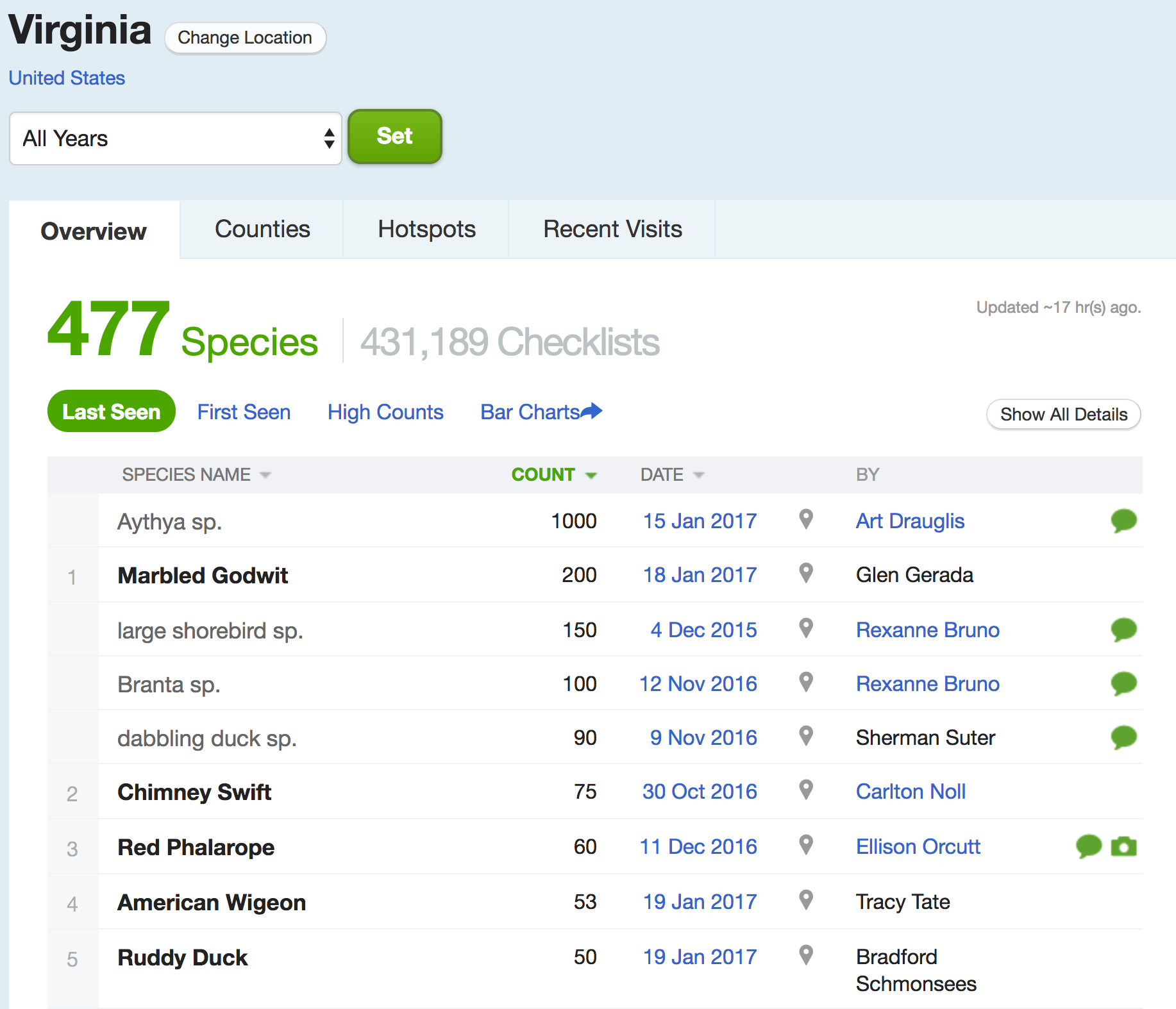}
 \caption{From www.ebird.org. A sample checklist for birds of Virginia. Some of the information included is bird count, location and field notes.}
  \label{fig:checklist}
 \vspace*{0.1in}
\end{figure}

Starting from a particular region such as North America for example, a system backend should be maintained containing data downloaded from Flickr and ebird containing bird sighting information, location, photographs, counts, habits, season etc. For each bird species an informed choice (based on experimentation) of classifiers such as Support Vector Machine or logistic regression or neural network should be made. A choice between the first two(linear) or the latter(non-linear) classifiers can be made based on the size of the data for each species and the number of features. The features here include for example the time of the sighting (the day could be classified into multiple periods containing a few hours with each forming a different mutually exclusive feature), the season of the sighting (each season forming a different feature) the sighting count for that species (can be classified as high and low). Other features imperative for impacting the likelihood of sighting a species should be extracted from the interview sessions with experts. Each bird sighting data can be represented as one-hot vector with 1 if a particular attribute is present or 0 if it is absent. Such representations are common in representing text documents \cite{Pang:2002:TUS:1118693.1118704}. Now the above mentioned classifiers can be trained for each species to predict the probability of sighting a bird from that species.

\begin{figure}[ht!]
 \centering
  \includegraphics[width=0.5\textwidth]{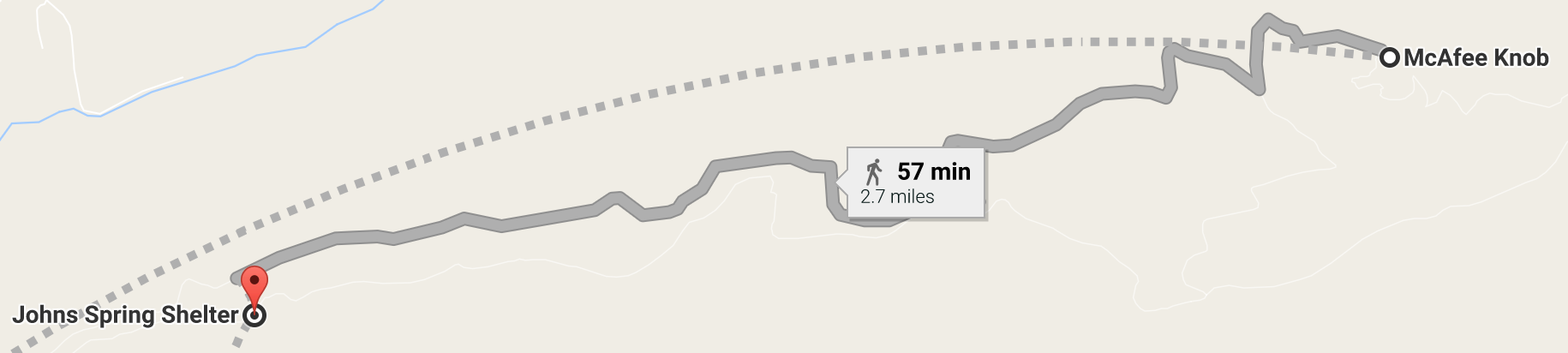}
 \caption{Johns Springs shelter to McAfee's knob --- a part of the Appalachian trail.}
 \label{fig:trail}
 \vspace*{0.1in}
\end{figure}

\begin{figure}[ht!]
  \includegraphics[width=0.25\textwidth]{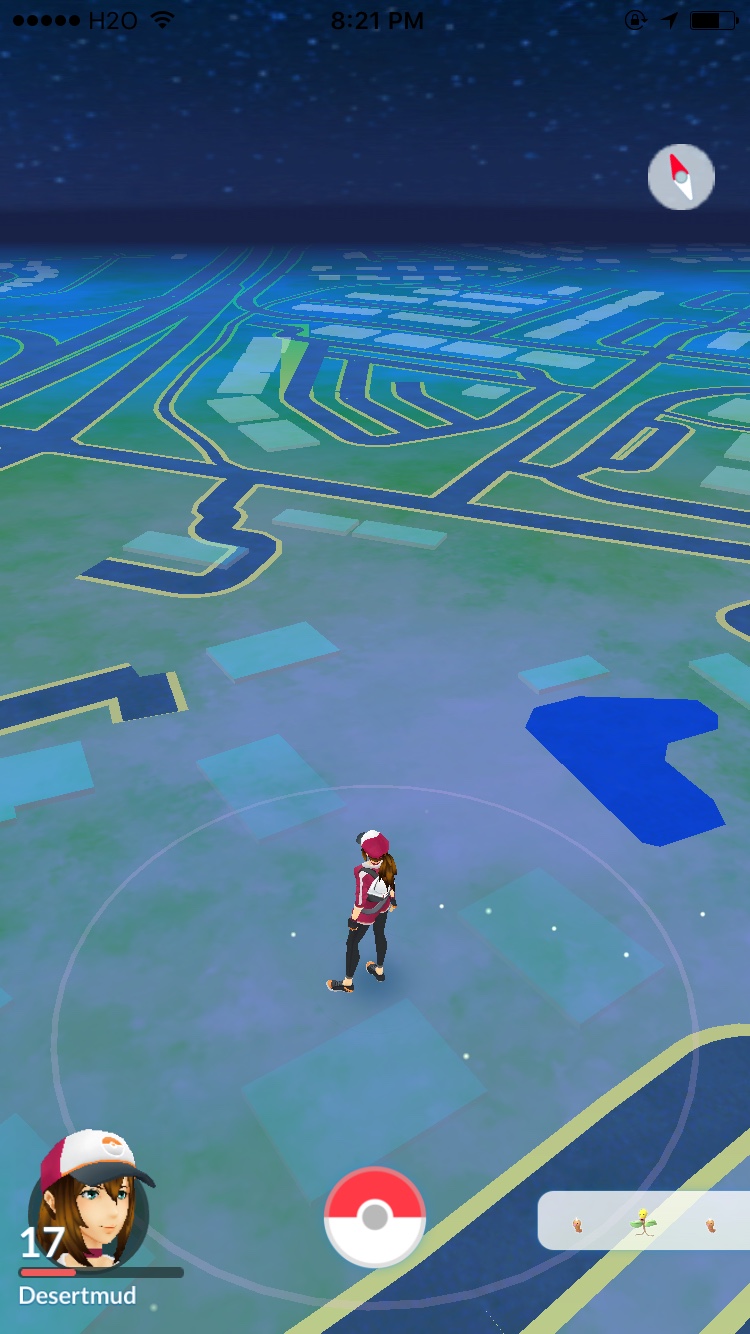}\includegraphics[width=0.25\textwidth]{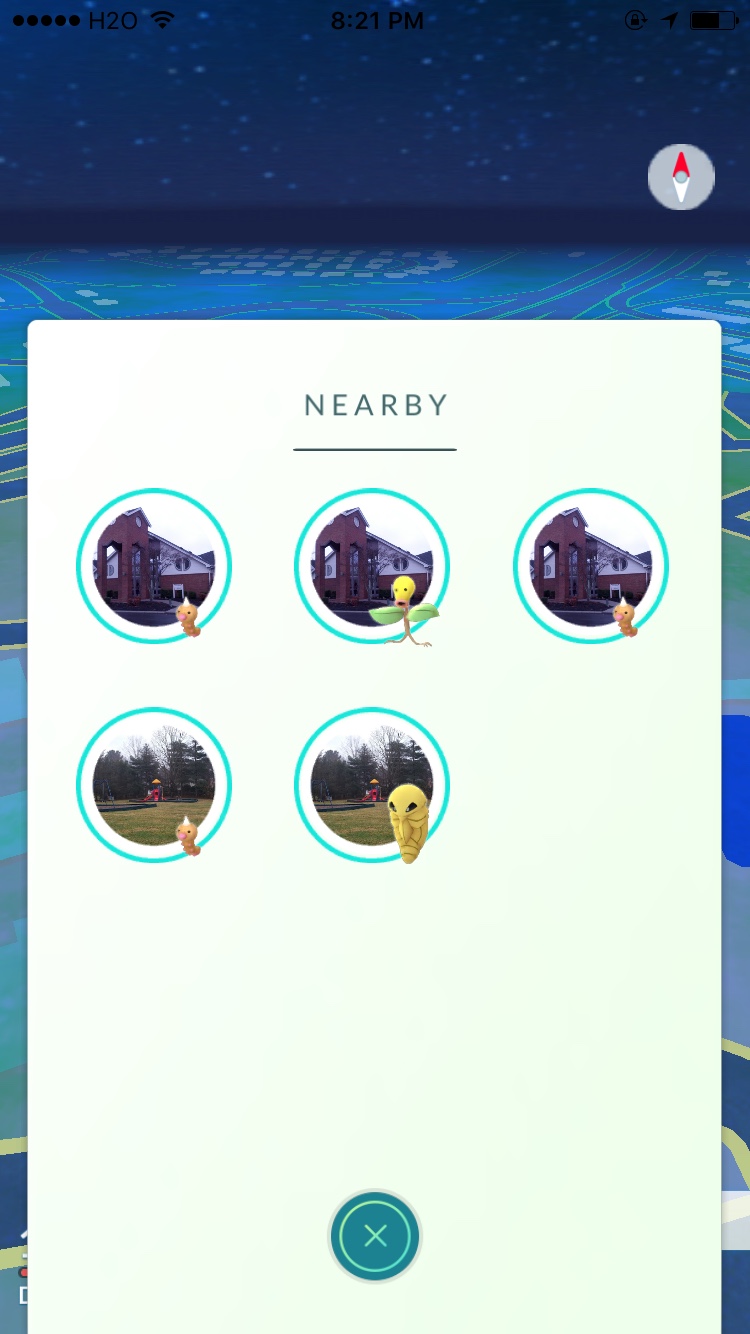}
 \caption{For displaying the map of the hike a possible interface choice could be a 3D world map(left) and an interface similar to ``nearby Pokemon'' (right) could work well for showing nearby birds.}
  \label{fig:pokemon}
 \vspace*{0.1in}
\end{figure}

\begin{figure}[ht!]
 \centering
  \includegraphics[width=0.5\textwidth]{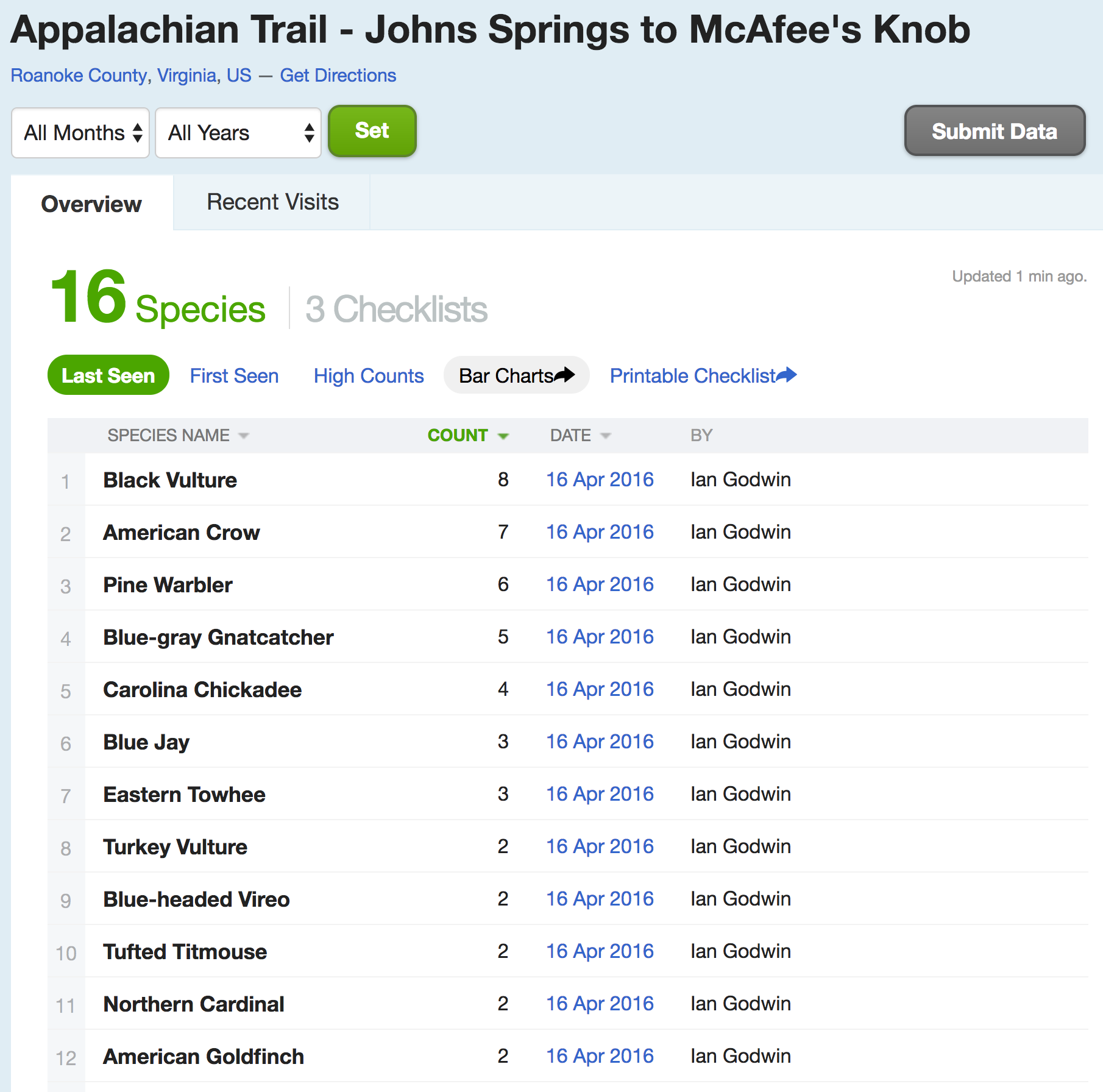}
 \caption{From www.ebird.org. Information aggregated from checklists submitted for the Johns Springs to McAfee's Knob trail.}
 \label{fig:trail-checklist}
 \vspace*{0.1in}
\end{figure}

Consider the 4.5 km trail from Johns Springs shelter to McAfee's knob --- a part of the Appalachian trail (Figure \ref{fig:trail}). Figure \ref{fig:trail-checklist} shows the aggregated information of bird checklists submitted on ebird where Black Vulture, American Crow, Pine Warbler and Blue-gray Gnatcatcher can be seen as frequently spotted birds.  From the current user conditions such as the location of the user, time of the day, season etc. a one hot vector can be formed and for each of the top 10(say) frequently sighted birds probability of their sightings can be predicted using the trained classifiers for those species. Based on the probability StB would prompt the user for the birds most likely to be sighted. The interface for displaying the nearby birds should possibly be similar to game of Pokemon Go(Figure \ref{fig:pokemon}).

Another important design consideration is displaying the map of the trail route.
In \cite{Kray:2003:PRI:604045.604066} Krey et al. perform a comparative study of different visualizations on mobile devices for presenting directional information while wayfinding outdoors. They compare text instructions, speech instructions, 2D maps and 3D maps. As per their findings many users preferred 3D maps and thought they were 'more fun' to use. They found that 3D visualizations are well suited for situations where time and technical resources are not an issue and where the available positional information is somewhat imprecise: the realistic presentations allows the user to search her environment visually for specific objects, and
then to align herself accordingly, thereby compensating the imprecision. This combination of realism combined with medium cognitive load on the user would make 3D or 2.5D maps a good choice for an app like StB (Figure \ref{fig:pokemon}). My choice for the visualization is also reinforced by the increasing popularity of realistic map visualizations in wildly popular applications like Uber and Google Maps for example.

When the user sees nearby birds on the app, she can be alert for those birds until they're still in the vicinity. To gain points the user has to click a photograph of the bird in which the bird is clearly visible and submit to the system. The app provides information about the bird such as it's call, photographs, nest photos, similar looking species, habits and other high level information. The app is designed for novice birders or non-birders so it is important to not overload the users with detailed information as present in some of the advanced bird guides. As discussed previously interview with 1-2 ornithologists will help shed light on what information can be included and what should be excluded for beginners.

\subsection{Verification}
Spotting the bird using the information provided in the app and photographing it are both relatively difficult tasks especially for novice birders. It is important to click a picture such that the bird is clearly visible in the picture. This part can be tricky as I have observed from my own field experiences bird watching. Once a picture is clicked and uploaded the system uses a combination of human input and computer vision to figure out if the bird is indeed what the user thinks it is. Given a possibly low quality photo (typical camera resolution on iPhone/iPad devices is 8 megapixels) and candidate bird pictures, the system has to figure out if the bird is indeed a picture of the candidate. 

In Visipedia Circa \cite{Belongie201615} Belongie et al. propose the idea of a Wikipedia visual counterpart -- a system for discovering and organizing visual information and making it easily accessible to anyone. The system  with a bigger role for automation --- a decentralized, continually improving collaborative network of people and machines. As one instantiation of Visipedia, authors develop Merlin bird ID, an app that uses hybrid human and computer vision algorithms \cite{Branson:2010:VRH:1888089.1888123} \cite{Branson2014} for identifying bird species. Merlin bird ID is made as a field guide for bird identification where as StB is a game. The technology behind Merlin Photo ID combines the skills of ignorant humans with poor-sighted machines and achieves quick and accurate bird identification. In StB the problem is similar to that of fine-grained recognition. It is different from fine-grained recognition in that we want to verify if the bird in the photo belongs to the candidate bird as opposed to recognizing the bird species for scratch.
 
One possible approach for confirming if the clicked photo indeed belongs to the candidate bird is to ask crowdworkers on Amazon Mechanical Turk(AMT) \cite{Kittur:2008:CUS:1357054.1357127} and verify the species by using one of the many crowd workflows developed for class verification\cite{Karger:2013:ECM:2494232.2465761}. But unfortunately a recent study revealed \cite{Horn2015} that the fine-grained categories in CUB-200 \cite{WelinderEtal2010} and ImageNet \cite{NIPS2012_4824}, both of which used AMT to clean the datasets, have significant type I and type II errors.
In StB we need accurate information if the bird photographed by the user is indeed the candidate so we cannot rely on the approximate answers by novice crowdworkers. This puts the crowd-alone approach in the backseat.

\begin{figure}[ht!]
 \centering
  \includegraphics[width=0.5\textwidth]{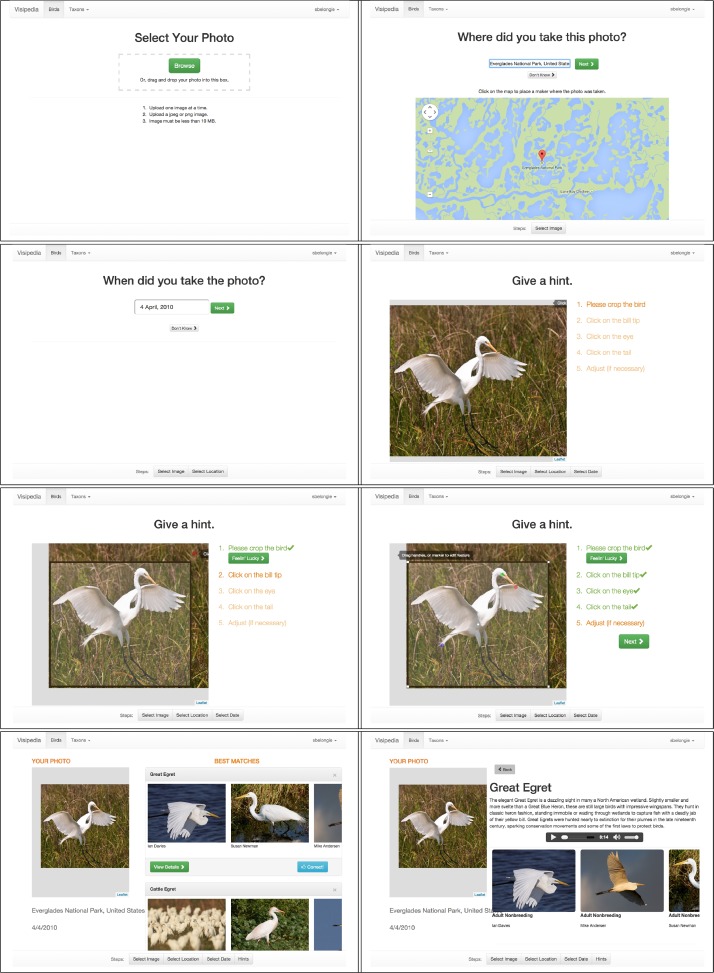}
 \caption{(From [2]) Use of Merlin photo-id application (http://merlin.allaboutbirds.org/photo-id). The user first uploads a picture (the photograph used in this example was taken from http://parkorbird.flickr.com), then provides the system with the location and date where the photo was taken (``don't know'' is a legal input). Finally, the user may click on three landmarks (bill tip, eye and tail end) to help the system locate the bird. The system outputs a list of likely bird species, prioritized by probability. The user may then access additional information by clicking on live links provided with the output.}
 \label{fig:merlin}
 \vspace*{0.1in}
\end{figure}

In this case there is no need to perform fine-grained recognition from scratch, only fine-grained verification. This makes me think if it is possible to use pure visual similarity based approaches \cite{Lew:2006:CMI:1126004.1126005} to match the bird photos. Ofcourse to understand how effective this is experiments with real data need to be performed. Another way would be to adopt a hybrid approach as described in \cite{Branson2014} \cite{Branson2010}. In the hybrid approach the system intelligently asks questions to the users. Users answer those questions by clicking on the bird photo which helps the system resolve possible conflicts between similar looking species. Figure \ref{fig:merlin} illustrates this.

If the user clicked picture matches the species she scores points. Crossing each level needs accumulation of a certain number of points. As the user goes to higher levels system starts suggesting rarer and more difficult to sight birds. 
 
\section{Evaluation}
In lab evaluations and in the wild evaluations can be quite different. In the wild studies show how people come to appropriate technology on their own terms in their own time. \cite{Rogers:2011:IDG:1978822.1978834} authors outline competing theories for interaction in the wild such as theory of embodiment or felt experience which emphasizes the whole experience of
a technology in terms of its interconnected aspects, rather than its fragmented aspects (for example,
its usability or utility) \cite{McCarthy:2004:TE:1015530.1015549}. In this case in the wild studies make the most sense. My first evaluation goal is to assess if the users learned the names of the bird species in their surroundings after using the app. For this reason I propose to conduct a study with non-birders and novices over a pre-decided route and get a list of all birds on that route from the backend. In this study the route will be fed into the user's device, one of the researchers will accompany the user on the hike and take observation notes. Observation notes will include information such as birds sighted by the user, birds missed, birds photographed, birds identified without help from the app, birds identified by call etc. After the hike the novice user will be given a test based on the birds encountered by the user on the hike(from the observation notes). The test will exclude the birds that user identified without the help of the app, since it is possible that the user already has knowledge about those birds. In the test the user will be asked to identify birds based on their photo and sound. The performance on this test will demonstrate to effectiveness of the app. 

Besides evaluating for the main goal specified above  I will also perform evaluations for usability of the app. In \cite{Bonsignore:2013:SSL:2491500.2491506} authors perform a longitudinal multimethod study collecting qualitative as well quantitative data and incorporating feedback into app design. I propose to adopt a similar approach to evaluate StB. I will collect qualitative data describing subjective experiences of the users in using the app. Quantitative data can be collected in the form of GPS trace of the user. This will give us the information about distance traveled by the user in pursuit of birds and how off is the distance from the trail length. Information about the number of birds missed can be calculated by subtracting the number of birds spotted from the number of birds expected to be sighted on that trail. Although this information is reliable only within a certain confidence interval since likelihood of sighting a particular bird is only a probability. More accurate misses vs captures can be calculated by sending an expert along with the users to record all the sightings separately.
\section{Discussion}
One of the important implications of an app like this would be increased physical activity as observed in similar games \cite{info:doi/10.2196/jmir.6759}. Gamifying the process accomplishes the dual goals of first pursuading the non-active users to go out there in the wild in the pretext of playing a game and engaging the casual users for longer hours increasing their physical activity \cite{info:doi/10.2196/jmir.6759} When the game involves collectibles there is an added motivation, collectibles in this context meaning the number of bird species seen and photographed by the user. On the other hand it can be argued that such an app would increase the cognitive load of a user looking for a casual hike leading to less engagement and use. Some naturists can also argue that an application like this would lead to increased human intrusion in the natural world. Another implication would be the spirit of travel that such an app would foster with the motivation of discovering local bird species in other parts of the world.

There are some difficult technological questions to be addressed. For instance getting enough data for a number of bird species to be able to train reliable classifiers is a challenging task. Another important question is if the classification should run on the device(client) or on the server, there would be issues of latency in the latter case and memory in the former. For fine-grained verification even Merlin photo-id is able to recognize only 650 of common North American bird species. So if the verification pipeline is to be based on Merlin dataset, we would be limited by the number of species and the locales where the app would be of use.

Besides that there is an inherent limitation of weak GPS signal in forest areas. This issue can be addressed by offline maps and suggestions.

\section{Conclusion and Future Work}
In this paper I have proposed the idea of a bird spotting game. I have presented the design and technological considerations to be taken into account while developing such an app. Finally I also discuss evaluation mechanisms to evaluate this app. A possible direction for future work would be to give a social aspect to the game in which the users can communicate about their sightings with each other and share information with each other in real time. Designing the social network around this app would be yet another challenge because special care would need to be taken to not intrude into the wilderness and scare away the birds.

\section{Acknowledgments}
I would like to thank the Computer Science department of Virginia Tech and HCI qualifier committee 2017 for giving me the opportunity to work on this paper.

\bibliographystyle{SIGCHI-Reference-Format}
\bibliography{sample}

\end{document}